\def\lapp{\mathrel{\mathop{<}\limits_{{}^\sim}}}

\documentstyle[12pt]{article}

\newmathalphabet{\bfmath}
\newmathalphabet{\rmmath}
\newmathalphabet{\itmath}
\addtoversion{normal}{\bfmath}{cmr}{bx}{n}
\addtoversion{normal}{\rmmath}{cmr}{m}{n}
\addtoversion{normal}{\itmath}{cmr}{m}{it}

\title{ The Charge Response of \\
a Meson-Correlated Relativistic Fermi Gas
\thanks{This work is supported in part by
funds provided by the U.S. Department of Energy (D.O.E.) under cooperative
agreement \#DE-FC02-94ER40818    .}
}
\author{{M.B. Barbaro}$^a$, {A. De Pace}$^a$,
{T.W. Donnelly}$^b$ and A. Molinari$^{a}$
\\ \\
\em $^a$ Dipartimento di Fisica Teorica dell'Universit\`a
 di Torino and \\
\em Istituto Nazionale di Fisica Nucleare, Sezione di Torino, \\
\em via P.Giuria 1, I-10125 Torino, Italy \\
\em $^b$ Center for Theoretical Physics, \\
\em Laboratory for Nuclear Science and Department of Physics, \\
\em Massachusetts Institute of Technology, Cambridge, MA 02139, USA}

\date{}

\begin{document}

\maketitle

\begin{abstract}
The quasielastic longitudinal electromagnetic response function $R_L$ is
studied within the context of a model that extends our previous investigations
of pionic correlations and currents.  Four mesons are now employed
($\pi$, $\rho$, $\sigma$ and $\omega$, via the Bonn potential) and
the many-body dynamics are extended to the full random-phase
approximation built upon a Hartree-Fock basis.  Wherever possible the
Lorentz covariance of the problem is respected.  The first three
energy-weighted moments of the reduced response are computed, namely,
the zeroth moment (Coulomb sum rule), the first moment (related to the
position of the quasielastic peak) and the second moment (related to
the peak width or variance).  We discuss how with a modest downward adjustment
of the Fermi momentum it is possible to obtain the expected zeroth and second
moments; this implies that the nuclear system has a lower density than
that required when using the free relativistic Fermi gas and accordingly
that the interaction effects are weaker than one might initially find.
\end{abstract}

\vfill

CTP\#2399 \hfill January 1995

\eject

\section{Introduction}
\label{sec:intro}

In past work \cite{Alb90,Alb93} we have addressed the problem of setting up
both the {\em longitudinal} ($R_L$) and the {\em transverse} ($R_T$) responses
that enter in the expression of the inclusive cross section for inelastic
scattering of electrons from nuclei. The inelasticity domain of concern here is
that of the quasielastic peak (QEP).
Our studies were based on the relativistic Fermi
gas (RFG) model and the guiding principles of our investigation
were relativistic covariance and global gauge invariance (fulfillment of the
continuity equation) --- indeed much effort was expended in adhering to these
two fundamental constraints. In particular, the choice of the RFG model was
prompted not only by the assumption that the nuclear surface is likely of
secondary importance in very inelastic nuclear phenomena, but also by the
recognition that the two principles referred to here are easier to implement
in a translationally invariant system than in a finite nucleus.

Moreover, in our previous work we resorted to the
Hartree-Fock (HF) and antisymmetrized random phase approximation
(RPA) many-body frameworks, {\em adding} their contributions to get the nuclear
responses. These procedures are correct when the treatment is limited to
first-order contributions in a perturbative expansion, an approach that has
some merit when the emphasis is being placed on achieving a consistent
treatment of the forces and currents (the meson-exchange currents (MEC) are
in fact usually expressed through first-order perturbative diagrams) and
when the interaction between
nucleons is carried solely by the {\em pion} whose associated correlations are
weak enough to make a first-order treatment quite reliable.
Indeed, comparison with the infinite-order calculations showed this to be
the case, at least for momentum transfers higher than roughly 300~MeV/c
\cite{Alb93}.

In the present work we extend the previous investigations in two
directions by
\begin{itemize}
\item[i)]{including, beyond the $\pi$, additional mesons (namely the $\rho$,
$\sigma$ and $\omega$) in the description of the nuclear force, and}
\item[ii)]{calculating the nuclear electromagnetic responses in the fully
antisymmetrized RPA scheme {\em evaluated in a HF basis}.}
\end{itemize}
Clearly, item ii) is implied by i), since the forces carried by the $\rho$,
$\sigma$ and $\omega$ are too strong to be dealt with in first order.
On the other hand, the requirement of global gauge invariance is now much
harder to implement in a consistent manner and we leave this issue to be dealt
with in a future paper.

Here we address the specific problem of finding how the forces
between the nucleons shape the quasielastic response in the range of momenta
between 300 MeV/c and 1 GeV/c, confining ourselves to the charge response
$R_L$ (the transverse one, $R_T$, is treated in a companion paper).
To assess the impact of the forces on $R_L$
we focus not only on the longitudinal response itself, but on three
derived quantities as well, namely the sum rule, the mean
energy and the variance of the response (we shall later provide
precise definitions for these quantities). Note that these observables
relate only to the many-body
content of $R_L$, which should therefore be carefully disentangled from
nucleonic aspects of the problem as we shall see in detail.

In performing the present investigation an underlying theme has been the
role played by the pion.
Apart from the self-evident importance of the pion in the nuclear dynamics, our
interest has also been prompted by the findings of ref.~\cite{Bar94}, where it
has been shown that, owing to its isovector nature, the pion acting alone
induces a remarkable enhancement of the longitudinal vector weak neutral
current
response. This in turn entails a huge effect on a certain observable related to
the asymmetry as measured in parity-violating inclusive electron scattering.
Our wish to explore whether or not pionic effects survive when other powerful
components of the nucleon-nucleon interaction are switched on is therefore
understandable.

Significantly, we have found that the role of the pion, far from being
``suppressed'', is actually ``enhanced'' by the shorter-range part of the
nuclear force through an interference mechanism hidden in the complexities of
the charge response of the correlated RFG.
On the other hand, it is well-known that pionic effects in the ground-state
energy of a spin-isospin saturated system like the RFG or a closed-shell
nucleus
are suppressed, because the key ingredient, namely the Hartree contribution, is
eliminated by the spin-isospin traces.

It thus appears that in nuclear matter the ground state and the QEP involve
different aspects of the dynamics, namely that the action of the pion
in the ground state is hampered, while in the QEP it is enhanced.
This is a long-standing question that does not appear to have received
unambiguous experimental resolution; however,
a recent reanalysis \cite{Tad94} of polarized ($\vec{p},\vec{n}$)
experiments at LAMPF shows that the data are compatible with a softening
of the spin-longitudinal response induced by the pion, as predicted many years
ago \cite{Alb82}.

Since the developments and results presented in the body of this work are
quite involved, in the following paragraphs we summarize some of the
salient findings before entering into those more detailed discussions.
In sects.~\ref{sec:HF} and \ref{sec:HF-responses} we build
the HF mean field utilizing as input the $\pi$, $\rho$, $\sigma$ and $\omega$
pieces of the well-known relativistic Bonn
potential \cite{Mac87}. We show that the HF field is
strong and induces a very substantial hardening of $R_L$.
This arises mainly from the tadpole diagram via the cooperative action of the
$\sigma$ and the $\omega$, with however an additional sizable contribution
having the same sign and stemming from the $\rho$ meson through the ``oyster''
(Fock) diagram. Worth noticing is that the momentum dependence of the tadpole
diagram, and hence the hardening of the charge response, is of relativistic
origin.

As we shall see, we are able to account quite accurately for the action of
the HF field on the nuclear response simply by shifting the energy
$\omega$ transferred to the nucleus by an amount $\Delta\omega$ and by changing
the bare nucleon mass $m_N$ into a dressed one, $m_N^*$.
It turns out that $\Delta\omega$ and $m_N^*$ depend upon the momentum transfer
$q$ and the Fermi momentum $k_F$.
This result is the key for obtaining RPA results in a HF basis
without having to resort to quite heavy computations.
The strength of the mean field up to the largest momentum considered here,
$q=1$ GeV/c, is also reflected by the variance of the HF
longitudinal
response (essentially the half-maximum width) which is significantly larger
than that occurring in the RFG model, but for the lowest momenta.

In sect.~\ref{sec:RPA} we investigate the roles played by the ring
and exchange correlations in $R_L$. We deal exactly with the
rings, whereas the exchange diagrams are treated in the framework of the
continuous fraction method truncated at the first iteration. This scheme
actually provides exact results for zero-range forces and also turns out
to be quite accurate for finite-range interactions as long as the
momentum transfer is not too small (say $q\geq 300$ MeV/c)

In the first part of sect.~\ref{sec:RPA} we consider only diagrams going
forward in time (antisymmetrized Tamm-Dancoff (TD) approximation) and, to
assess the impact of the mean field on the TD results better, these are
presented
with fermions propagating both freely and in the presence of a HF field.
In the TD scheme (and as well in RPA) a struggle among the different
contributions to the charge response takes place.
Specifically the battle arises from the interplay between the direct (ring)
and exchange diagrams and among the various mesons carrying the force.
To unlock this complex situation it might help to recall that
\begin{itemize}
\item[a)] {the momentum behaviour of the ring and exchange diagrams is rather
similar, both of them decreasing with $q$ and becoming much reduced for
$q>500$ MeV/c;}
\item[b)] {generally, the light (heavy) mesons are more effective in
transmitting the interaction at small (large) $q$, providing of course that
the coupling constants are large enough.}
\end{itemize}

Thus, considering first the ring diagrams, it is found that they indeed
contribute sizably up to quite large transferred momenta and are obviously
supported only by the $\sigma$ and $\omega$ mesons.
These dramatically reshape the free charge response in the isoscalar
channel when acting alone, leading to the occurrence of collective modes lying
above (the $\omega$) and below (the $\sigma$) the energy range of the QEP
that carry a large fraction of the charge strength (hence leading to a huge
depletion of $R_L$ inside the QEP).
However, when acting together, the $\sigma$ and the $\omega$ interfere so
strongly that the ``ring'' collective modes disappear and the only remnant of
the $\sigma$--$\omega$ interplay is a still appreciable damping
(but for the lowest energies) and flattening at 300 MeV/c (but not for larger
momenta) of $R_L$. In addition a {\em softening} of $R_L$ takes place ---
this is large at 300 MeV/c, still significant at 500 MeV/c and then fades
away at higher momentum transfer
in accord with a). On the other hand, in accord with b) the
$\sigma$ wins over the $\omega$, as it also does for the ground-state binding
energy.

Turning to the exchange, it is contributed to by all of the mesons.
Furthermore it opposes the action
of the rings, thus leading in accord with b) to a hardening of $R_L$,
which shows up as a dramatic peak in the response at $q=300$ MeV/c,
with some
shadow of it remaining even at 500 MeV/c. This peak corresponds to an
``exchange'' collective mode embedded in the particle-hole ($ph$) continuum
that arises from the repeated exchange of the attractive $\pi, \rho$ and
$\sigma$ mesons, which are more powerful than the $\omega$ at these momenta
inside a $ph$ ring. Of considerable significance in this
connection is the role of the pion and, to a
much lesser extent, of the rho: indeed, {\em without the pion} the exchange
peak
would {\em not} appear at all!

Actually, when the ring and the exchange act together, the repulsive peak
disappears, although some hardening of $R_L$ remains at 300 and 500 MeV/c,
thus showing the dominance of the exchange correlations over
those from the ring diagrams. At larger momenta the ring and exchange
diagrams, beyond being quite small, almost completely cancel each other.

Summarizing: the TD collective effects in the charge channel arise
basically from the exchange diagrams.
They show up essentially through a hardening of the charge response at
intermediate momenta (due to the attractive $\sigma$, but particularly to the
$\pi$), which persists up to about 500 MeV/c.  However, as we shall see, the
TD correlations appear to be somewhat overcome by the HF field:
thus the TD longitudinal response evaluated in a HF basis, while being
appreciably depleted at intermediate momenta, is actually shifted at higher
frequencies for all of the transferred momenta considered by an amount which is
rather constant but for the lowest momenta.

In the second part of sect.~\ref{sec:RPA} the diagrams going backward in time
are explored (antisymmetrized RPA). The ground-state correlations
(gsc) thus introduced appear to strengthen the
action of the attractive ring diagrams somewhat, while at the same time
reducing the contribution of the repulsive exchange.
Indeed the ``exchange peak'' at $q=300$ MeV/c, referred to above in the TD
scheme,
now no longer shows up and is instead replaced by a ``ring peak'', associated
with the $\sigma$ meson and appearing in the direct channel.
Yet the non-linearity of the RPA is such that, even in the presence of gsc,
the exchange term still prevails over the direct one when they are combined
together and what remains is actually an appreciable hardening.
In addition to this, the RPA correlations at momenta up to about 500 MeV/c
(and likewise the TD ones,
but more markedly so) have a tendency to quench and flatten $R_L$, particularly
when evaluated in a HF basis, namely they produce trends that appear to be
supported by some experimental evidence \cite{Bar83}-\cite{Bla86}.
At larger momenta, the contribution of the gsc essentially vanishes, as can
also be inferred from the observation that here the sum rule is well fulfilled.

The important role that the pion also plays in the RPA framework should be
emphasized. Indeed, without it the {\em ring} peak would survive the
confrontation with the exchange contribution: accordingly, the RPA response in
a
free basis, and in a HF one as well, would still display a peak on the
low-energy side of the response.
This would be entirely out of touch with reality, if
only the $\sigma$ and the $\omega$ were to act, and still there in a scheme
encompassing the $\rho$ as well. Thus, the statement that the RPA charge
response requires the pion appears to be justified.

In summary, the RPA correlations are found to yield an $R_L$ that is
appreciably less hardened than the one obtained in the TD scheme,
but somewhat more quenched and with a slightly larger variance.
This finding is in accord with the general result that the gsc lower (enhance)
the response when the latter is hardened (softened) in order to comply with the
energy-weighted sum rule. Finally, as in TD, the HF mean field still tends to
play a prominent role within the RPA and again the role of the pion is crucial.

{}From our analysis of the longitudinal response based on the HF--TD and
HF--RPA schemes with the Bonn potential as an input, we thus conclude that the
correlations' effects are very substantial and to a large extent appear
to disrupt the RFG predictions. One is accordingly led to address the
question whether or not part of the interaction
might be accounted for by ``renormalizing'' the only parameter
characterizing our many-body system, namely the Fermi momentum $k_F$.
Our choice, namely $k_F=225$ MeV/c, has been dictated by the analysis of the
$^{12}$C data where such a value turns out approximately to account over
a range of momenta
for the width of the longitudinal response when the latter is analyzed in
terms of the RFG model.
Since the width at half-maximum together with the area and position of the
QEP are the central features of the inclusive
charge response (or, in general, of {\em any} response function),
we have searched within the scheme of our meson-correlated
RFG for a $k_F$ capable of accounting for the experimental half-width or,
which is roughly the same thing, of yielding the same half-width as the RFG
with
$k_F=225$ MeV/c. In other words we have applied what can be defined as
a principle of the {\em invariance of the variance}, the latter being of
course related to the half-width of the response. The integrated response
and peak position are then computed using this renormalized model.

The results are discussed in sect.~\ref{sec:variance}.
We have been successful in adhering to this principle,
providing we let $k_F$ vary slightly with $q$. The values of $k_F$
necessary for achieving this
turn out always to be lower than 225 MeV/c by about 10\% and to be gently
increasing with
$q$ ({\it i.~e.,\/} the larger the value of $q$, the smaller the required
renormalization of the Fermi
momentum). Because of the non-linear dependence on $k_F$ of the interaction
effects in our model,
the impact of the correlations on $R_L$ is overall found to be appreciably
diminished.
Note that, due to the different ranges of the forces carried by the different
quanta,
the heavier mesons are those most affected by the procedure outlined above,
whereas the pion emerges almost unscathed from the renormalization of $k_F$.

Whether the results obtained in the {\em invariance of the variance} scheme
lead to closer contact with data or not is a question that remains to be
answered in detail. This we have not done
in the present work partly because
we have not yet completed the evaluation of the MEC contributions, which should
of course be added before attempting in-depth comparisons with the data,
but additionally because two
particle-two hole ($2p-2h$) excitations cannot be ignored in the QEP domain
and should be included as well in the analysis.

\section{The Hartree-Fock field}
\label{sec:HF}

Two diagrams contribute to the HF field, namely the direct (sometime referred
to
as {\em tadpole}) and the exchange (sometime referred to as {\em oyster})
diagrams displayed in Fig.~\ref{fig:se-diagrams}.
We have calculated their contribution to the nucleon self-energy utilizing the
Bonn potential, which is derived through a non-relativistic expansion of the
diagrams
corresponding to the exchanges of the $\pi$, $\rho$, $\sigma$ and $\omega$
mesons (actually the $\sigma$ meson should be viewed as a fictitious entity
corresponding to the exchange of a correlated pair of $S$-wave pions).
Here we have neglected the small $\eta$ and $\delta$ contributions in
ref.~\cite{Mac87} and for clarity retained only the largest pieces of the
potential.
It reads
\begin{equation}
V=V^\pi+V^\rho+V^\sigma+V^\omega,
\end{equation}
where
\begin{eqnarray}
V^\pi &=&
  (V_S^\pi\,\mbox{\boldmath $\sigma$}_1\cdot\mbox{\boldmath $\sigma$}_2 +
   V_T^\pi\,S_{12})\,\mbox{\boldmath $\tau$}_1\cdot\mbox{\boldmath $\tau$}_2
  \\
& & V_S^\pi(q) = V_T^\pi(q) =
  -\frac{g_\pi^2}{12m_N^2}\Gamma_\pi^2(q)\frac{q^2}{q^2+m_\pi^2},
  \nonumber \\
V^\rho &=& (V_0^\rho +
   V_S^\rho\,\mbox{\boldmath $\sigma$}_1\cdot\mbox{\boldmath $\sigma$}_2 +
   V_T^\rho\,S_{12})\,\mbox{\boldmath $\tau$}_1\cdot\mbox{\boldmath $\tau$}_2
  \\
& & V_0^\rho(q,P) = \left[g_\rho^2\left(1+\frac{3}{2}\frac{P^2}{m_N^2}
   \right)-\frac{g_\rho(g_\rho+4 f_\rho)}{8m_N^2}q^2\right]\Gamma_\rho^2(q)
   \frac{1}{q^2+m_\rho^2}
  \nonumber \\
& & V_S^\rho(q) = -2V_T^\rho(q) = -\frac{(g_\rho+f_\rho)^2}{6m_N^2}
   \Gamma_\rho^2(q)\frac{q^2}{q^2+m_\rho^2},
  \nonumber \\
V^\sigma &=& V_0^\sigma \\
& & V_0^\sigma(q,P) = g_\sigma^2\left[-1+\frac{1}{2}\frac{P^2}{m_N^2}
   -\frac{q^2}{8m_N^2}\right]\Gamma_\sigma^2(q)
   \frac{1}{q^2+m_\sigma^2},
  \nonumber \\
V^\omega &=& V_0^\omega +
   V_S^\omega\,\mbox{\boldmath $\sigma$}_1\cdot\mbox{\boldmath $\sigma$}_2 +
   V_T^\omega\,S_{12}
  \\
& & V_0^\omega(q,P) = g_\omega^2\left[1+\frac{3}{2}\frac{P^2}{m_N^2}
   -\frac{q^2}{8m_N^2}\right]\Gamma_\omega^2(q)
   \frac{1}{q^2+m_\omega^2},
  \nonumber \\
& & V_S^\omega(q) = -2V_T^\omega(q) = -\frac{g_\omega^2}{6m_N^2}
   \Gamma_\omega^2(q)\frac{q^2}{q^2+m_\omega^2}.
  \nonumber
\end{eqnarray}
In the above $S_{12}$ is the usual tensor operator and
\begin{equation}
\Gamma_\alpha(q) = \frac{\Lambda_\alpha^2-m_\alpha^2}{\Lambda_\alpha^2+q^2}
\end{equation}
the standard meson-nucleon form factor.
Moreover, $q\equiv|\bfmath{q}|$ is the momentum exchanged between the two
nucleons and $P\equiv|\bfmath{P}|$ is the average of their relative incoming
and outgoing momenta (as illustrated in Fig.~\ref{fig:NN-diagram},
$\bfmath{P}=(\bfmath{k}_1-\bfmath{k}_2+\bfmath{k}'_1-\bfmath{k}'_2)/4$).
The $\bfmath{P}$-dependent terms in the $\rho$, $\sigma$ and $\omega$
potentials
clearly correspond to a non-local interaction in coordinate space
and their important impact on the dynamics of the system will be discussed
later. We have omitted the spin-orbit terms
($\propto(\mbox{\boldmath $\sigma$}_1+\mbox{\boldmath $\sigma$}_2)\cdot
 \bfmath{q}\times\bfmath{P}$) because they do not contribute to the
responses in an infinite system.

For the convenience of the reader we report in Table~\ref{tab:Bonn} the quantum
numbers, the values of masses and cut-offs and the strengths of the couplings
for all of the mesons entering into the potential.
\begin{table}
\begin{center}
\begin{tabular}{|c|c|c|l|l|l|l|}
\hline
  meson  & $T$ & $J^\pi$ & $m$ (MeV/c) & $\Lambda$ (MeV) & $g^2/4\pi$ &
    $f^2/4\pi$ \\ \hline\hline
  $\pi$    & 1 & $0^-$ & 138.03      & 1300 & 14.9    & --            \\ \hline
  $\rho$   & 1 & $1^-$ & 769         & 1300 & 0.95
   & $0.95\times (6.1)^2$ \\ \hline
  $\sigma$ & 0 & $0^+$ & 550         & 2000 & 7.7823  & --            \\ \hline
  $\omega$ & 0 & $1^-$ & 782.6       & 1500 & 20      & --            \\ \hline
\end{tabular}
\end{center}
\caption{The quantum numbers, the masses, the cut-off parameters and the
coupling constants of the mesons included in the Bonn potential.
Note the two coupling constants characterizing the $\rho$.
  }
\label{tab:Bonn}
\end{table}

Note that in our approach, as in refs.~\cite{Alb90,Alb93}, the short-range
physics is again embedded in phenomenological
vertex form factors which cut off the nucleon-nucleon interaction in
a spatial region of size $\sim1/\Lambda$ around each nucleon. Whether this
is adequate or whether ladder diagram contributions to the responses should
be explicitly accounted for in the QEP domain is unclear.  This will need
to be addressed in future work.

In connection with the HF field, we recall that only the $\sigma$ and
$\omega$ contribute to the tadpole term. Introducing dimensionless parameters
\begin{equation}
\eta_F = \frac{k_F}{m_N}, \quad
\lambda_\alpha = \frac{\Lambda_\alpha}{m_N}, \quad
\mu_\alpha = \frac{m_\alpha}{m_N},
\end{equation}
and measuring both the momentum and the self-energy in units of the nucleon
mass
($\beta=k/m_N$,\,$\Sigma(k)=m_N\widetilde{\Sigma}(\beta)$),
the corresponding real and energy-independent self-energies read
\begin{eqnarray}
  \widetilde{\Sigma}_\sigma^H(\beta) &=&
    \frac{8}{3} \frac{g_\sigma^2}{4\pi^2\mu_\sigma^2}
    \left(\frac{\lambda_\sigma^2-\mu_\sigma^2}{\lambda_\sigma^2}\right)^2
    \eta_F^3\left(-1+\frac{3}{40}\eta_F^2+\frac{1}{8}\beta^2\right)
  \label{eq:se-s-h} \\
  \widetilde{\Sigma}_\omega^H(\beta) &=&
    \frac{8}{3} \frac{g_\omega^2}{4\pi^2\mu_\omega^2}
    \left(\frac{\lambda_\omega^2-\mu_\omega^2}{\lambda_\omega^2}\right)^2
    \eta_F^3\left( 1+\frac{9}{40}\eta_F^2+\frac{3}{8}\beta^2\right).
  \label{eq:se-o-h}
\end{eqnarray}
On the other hand all of the mesons contribute to the exchange ({\em oyster})
term and the associated self-energies, which are again real and
energy-independent, read
\begin{eqnarray}
  \widetilde{\Sigma}_\pi^F(\beta) &=&
    \frac{3}{8}\frac{g_\pi^2}{4\pi^2}
    \left[{\cal G}_\pi(\beta)+{\cal G}_\pi(-\beta)\right] \\
  \widetilde{\Sigma}_\rho^F(\beta) &=&
    -\frac{3}{2}\frac{g_\rho^2}{4\pi^2}
    \left[{\cal F}_\rho(\beta)+{\cal F}_\rho(-\beta)\right] \nonumber \\
    & & \quad +\frac{3}{16}
    \left(5\frac{g_\rho^2}{4\pi^2}+12\frac{f_\rho g_\rho}{4\pi^2} +
          4\frac{f_\rho^2}{4\pi^2}\right)
    \left[{\cal G}_\rho(\beta)+{\cal G}_\rho(-\beta)\right]
 \label{eq:sigmarho}\\
  \widetilde{\Sigma}_\sigma^F(\beta) &=&
    \frac{1}{2}\frac{g_\sigma^2}{4\pi^2}
    \left[{\cal F}_\sigma(\beta)+{\cal F}_\sigma(-\beta)\right] \nonumber \\
    & & \quad +\frac{5}{16}\frac{g_\sigma^2}{4\pi^2}
    \left[{\cal G}_\sigma(\beta)+{\cal G}_\sigma(-\beta)\right] \\
  \widetilde{\Sigma}_\omega^F(\beta) &=&
    -\frac{1}{2}\frac{g_\omega^2}{4\pi^2}
    \left[{\cal F}_\omega(\beta)+{\cal F}_\omega(-\beta)\right]
    \nonumber \\
    & & \quad +\frac{5}{16}\frac{g_\omega^2}{4\pi^2}
    \left[{\cal G}_\omega(\beta)+{\cal G}_\omega(-\beta)\right],
\label{eq:se-o-f}
\end{eqnarray}
where
\begin{eqnarray}
{\cal F}_\alpha(\beta) &=&
  \frac{\lambda_\alpha^2+\mu_\alpha^2}{\lambda_\alpha}
  \rmmath{tg}^{-1}\left(\frac{\eta_F+\beta}{\lambda_\alpha}\right) -
  2\mu_\alpha\rmmath{tg}^{-1}\left(\frac{\eta_F+\beta}{\mu_\alpha}\right)
  \nonumber \\
  & & \qquad
  + \frac{1}{2\beta}(\beta^2-\mu_\alpha^2-\eta_F^2)
  \rmmath{ln}
  \frac{(\eta_F+\beta)^2+\lambda_\alpha^2}{(\eta_F+\beta)^2+\mu_\alpha^2}
\end{eqnarray}
and
\begin{eqnarray}
{\cal G}_\alpha(\beta) &=&
  \lambda_\alpha(\lambda_\alpha^2-3\mu_\alpha^2)
  \rmmath{tg}^{-1}\left(\frac{\eta_F+\beta}{\lambda_\alpha}\right) +
  2\mu_\alpha^3\rmmath{tg}^{-1}\left(\frac{\eta_F+\beta}{\mu_\alpha}\right)
  \nonumber \\
  & & \qquad
  + \frac{1}{2\beta}(\eta_F^2\mu_\alpha^2-\beta^2\mu_\alpha^2+
    2\lambda_\alpha^2\mu_\alpha^2-\lambda_\alpha^4)
  \rmmath{ln}[(\eta_F+\beta)^2+\lambda_\alpha^2]
  \nonumber \\
  & & \qquad
  + \frac{\mu_\alpha^2}{2\beta}(\beta^2-\mu_\alpha^2-\eta_F^2)
  \rmmath{ln}[(\eta_F+\beta)^2+\mu_\alpha^2].
\end{eqnarray}

We thus see from the above that the Hartree field stemming from the action of
the $\sigma$ and of the $\omega$ mesons displays a term that is quadratic in
the momentum. The origin of this lies in the pieces of $V_0^{\sigma}$ and
$V_0^{\omega}$ that are proportional to $({q/2m_N})^2$ and $({P/2m_N})^2$,
namely, to the parameters characterizing the non-relativistic
expansion of the Bonn potential. In this connection note that the average
between the initial and final relative momenta of the interacting nucleons
in the Hartree diagram is
\begin{equation}
\bfmath{P} = {1\over 2} (\bfmath{k} - \bfmath{k}'),
\end{equation}
$\bfmath{k}$ being the momentum of the propagating nucleon (particle or hole)
and $\bfmath{k}'$ the momentum flowing in the bubble of
Fig.~\ref{fig:se-diagrams}.
Accordingly, in principle $\bfmath{P}$ is unbound. However, in practice,
in order that the truncation of the non-relativistic expansion of the Bonn
potential makes sense, both $|\bfmath{P}|$ and $|\bfmath{q}|$ should not
exceed,
roughly, the nucleon mass $m_N$.

Because of its quadratic dependence on the momentum the Hartree field
can be accounted for by an effective mass (of course, but for the constant
terms which enter into (\ref{eq:se-s-h}) and (\ref{eq:se-o-h})):
\begin{equation}
  m_N^* = {m_N \over \displaystyle 1 +
  {{\strut \rho} \over \displaystyle 4m_N^3} \left[
    {{\strut 3g_\omega^2}\over \displaystyle \mu_\omega^2}
      \left(
  {{\strut \lambda_\omega^2-\mu_\omega^2}\over \displaystyle
     \lambda_\omega^2}\right)^2+
    {{\strut g_\sigma^2}\over \displaystyle \mu_\sigma^2}
      \left(
  {{\strut \lambda_\sigma^2-\mu_\sigma^2}\over \displaystyle
     \lambda_\sigma^2}\right)^2
    \right]},
\end{equation}
where
\begin{equation}
\rho=\frac{2k_F^3}{3\pi^2}=m_N^3\frac{2\eta_F^3}{3\pi^2}
\end{equation}
is the uniform density of the Fermi system.
The effective mass is displayed as a function of $k_F$ in Fig.~\ref{fig:mstar},
where the individual contributions to $m_N^*$ arising from the $\sigma$ and
the $\omega$ are shown as well.
It is worth noting that the $\sigma$ and the $\omega$ cooperate in dressing
the bare mass of the nucleon whereas, as is well-known, they fight each
other in accounting for the binding energy of nuclear matter.
Clearly, this outcome reflects the {\em opposite} sign of the constant terms
and the {\em same} sign of the quadratic terms of the self-energies in
(\ref{eq:se-s-h}) and (\ref{eq:se-o-h}).

Turning to the Fock contribution, the nearly constant behaviour as a
function of the momentum of the combined Fock self-energy of the $\sigma$ and
$\omega$ (see Fig.~\ref{fig:se-fock}) is quite striking --- note that
a constant here produces no effect on the HF response of the system.
On the other hand the Fock self-energy of the $\rho$ is sizable: actually, if
acting alone, being repulsive, it would harden the charge response with respect
to the free case by about 20 MeV at $q=500$ MeV/c for $k_F=225$ MeV/c.
It clearly cannot be incorporated in a constant effective mass over a wide
range of momenta because, far from being quadratic, the Fock self-energy of the
$\rho$ displays a rather cumbersome momentum dependence (see
eq.~\ref{eq:sigmarho}).
Finally, we recall that in the $oyster$ diagram $P=0$.

In order to appreciate the dynamical impact of the HF field, let us now
calculate the longitudinal electromagnetic response in the HF approximation.
For this purpose we start by recalling its well-known expression
\cite{Alb80}, namely
\begin{eqnarray}
  R_L^{\rmmath{HF}}(q,\omega) &=& \frac{3\pi^2 A}{k_F^3}f_L^2(Q^2)
    \int\frac{d\bfmath{k}}{(2\pi)^3}
    \theta(k_F-k)\theta(|\bfmath{k}+\bfmath{q}|-k_F)
    \nonumber \\ & & \qquad\times
    \delta\left[\omega-\frac{q^2}{2m_N}-\frac{\bfmath{q}\cdot\bfmath{k}}{m_N}-
    \Sigma^{\rmmath{HF}}(|\bfmath{k}+\bfmath{q}|)
    +\Sigma^{\rmmath{HF}}(k)\right],
\label{eq:RLHF}
\end{eqnarray}
$A$ being the nuclear mass number (here and in the following we take
$N=Z=A/2$),
$Q$ the four momentum ($\omega,\bfmath{q}$), $f_L(Q^2)$ the longitudinal
$\gamma$NN vertex and $\Sigma^{\rmmath{HF}}$ the HF self-energy.
It is interesting that (\ref{eq:RLHF}) can be expressed as an integral over the
effective mass according to \cite{Alb80}
\begin{equation}
  R_L^{\rmmath{HF}}(q,\omega) =
    \frac{\xi_A}{8\pi^2}\frac{1}{\eta_F^3\kappa m_N^2}
    f_L^2(Q^2)\widetilde{R}_L^{\rmmath{HF}}(q,\omega),
\label{eq:RLHFbis}
\end{equation}
with $\kappa=q/2m_N$, $\xi_A=3\pi^2 A$ and
\begin{eqnarray}
  &&\!\!\!\widetilde{R}_L^{\rmmath{HF}}(q,\omega) =
    \left\{\theta(\eta_F-2\kappa)\int_{\eta_F-2\kappa}^{\eta_F}\!\!d\beta\,
      \beta \right.\nonumber \\ &&\qquad\qquad\qquad\left.
 + \theta(\eta_F-\kappa)\,\theta(2\kappa-\eta_F)
      \int_{2\kappa-\eta_F}^{\eta_F}\!\!d\beta\,\beta\right\}
      m_N^*(\sqrt{4\kappa^2+\beta^2+4\kappa y'}) \nonumber \\
    & & \!\!\!\!\!\!\!\!
 +\left\{\theta(\kappa-\eta_F)\int_{0}^{\eta_F}\!\!d\beta\,
      \beta + \theta(\eta_F-\kappa)\,\theta(2\kappa-\eta_F)
      \int_{0}^{2\kappa-\eta_F}\!\!d\beta\,\beta\right\}
      m_N^*(\sqrt{4\kappa^2+\beta^2+4\kappa y''}). \nonumber \\
\label{eq:HFtilde}
\end{eqnarray}
In the above
\begin{equation}
  m_N^*(\beta) = {m_N \over \left|\displaystyle 1 +
    {{\strut 1}\over\displaystyle\beta}
    {{\strut d\widetilde{\Sigma}^{\rmmath{HF}}(\beta)}
      \over\displaystyle d\beta}\right|}
\label{eq:eff-mass}
\end{equation}
is the momentum dependent effective mass of the nucleon, whereas $y'$ and $y''$
are roots of the equation
\begin{equation}
  \eta_F\psi-y-\frac{1}{2\kappa}
    [\widetilde{\Sigma}^{\rmmath{HF}}(\sqrt{4\kappa^2+\beta^2+4\kappa y})-
     \widetilde{\Sigma}^{\rmmath{HF}}(\beta)]=0,
\label{eq:se-equation}
\end{equation}
satisfying the constraints
\begin{equation}
  (\eta_F^2-4\kappa^2-\beta^2)/4\kappa\le y'\le\beta
  \qquad\mbox{and}\qquad
  -\beta\le y''\le\beta.
\label{eq:se-constraint}
\end{equation}
It is of importance to realize, in connection with the basic result
(\ref{eq:HFtilde}), that in $R_L^{\rmmath{HF}}$
(and also in the transverse HF response)
only the {\em particle} (and not the {\em hole}) effective
mass explicitly appears.

Notice that the relativistic content of our approach is contained first of all
in the relativistic scaling variable of the Fermi gas $\psi$.
For the latter we use, rather that the exact one \cite{Alb88}, the approximate
but quite accurate relativistic expression
\begin{equation}
  \psi = \frac{1}{\eta_F}\left[\frac{\lambda(\lambda+1)}{\kappa}-\kappa\right],
\label{eq:psi-rel}
\end{equation}
with $\lambda=\omega/2m_N$.
Next, relativity is also embedded in the longitudinal $\gamma$NN
vertex. For this we again employ the approximate but almost exact
\cite{Alb90} expression
\begin{eqnarray}
  f_L^2(Q^2) &=& (1+2\lambda)\left\{\frac{1}{1+\tau}[G_E^{m_t}(\tau)]^2+
    \tau[G_M^{m_t}(\tau)]^2\frac{\eta_F^2}{2}
    (1-\psi^2)\right\}
\label{eq:fL}
\end{eqnarray}
$G_{E,M}^{m_t}$ being the proton and neutron (with isospin
projections $m_t$) electric and magnetic Sachs form factors and
$\tau=|Q^2|/4m_N^2$.

Before displaying the results for $R_L^{\rmmath{HF}}$, we now briefly
outline an approach that greatly simplifies the RPA calculation of the nuclear
responses based on a HF (rather than on a free) basis.
This approach stems from the observation that the total HF nucleon self-energy
induced by the $\pi$, $\rho$, $\sigma$ and $\omega$ mesons, while not
simply parabolic
in the momentum, nevertheless lends itself to an interpolation with {\em two}
parabolas, one below (hole states) and one above (particle states) the Fermi
momentum. Accordingly, we set
\begin{equation}
  \widetilde{\Sigma}^{\rmmath{HF}}(\beta) = \bar{A}+\bar{B}\beta^2,
    \qquad\mbox{for}\;\beta\le\eta_F
\label{eq:fithole}
\end{equation}
and
\begin{equation}
  \widetilde{\Sigma}^{\rmmath{HF}}(\beta) = A+B\beta^2,
    \qquad\mbox{for}\;\beta>\eta_F,
\label{eq:fitparticle}
\end{equation}
$B$ and $\bar{B}$ being both positive.
However, while $\bar{A}$ and $\bar{B}$ are truly constant, $A$ and $B$ are in
fact momentum transfer dependent, since they are fixed by fitting the particle
self-energy over the range of momenta required for the calculation of a given
nuclear response. This range, of course, varies with $q$. As already noted,
the procedure outlined here allows a quite accurate fit of the self-energy
entering into the charge response.

In Table~\ref{tab:bipar} we give the values of the parameters appearing in
(\ref{eq:fithole}) and (\ref{eq:fitparticle}) for $q=300,$ 500, 800 and
1000 MeV/c.
\begin{table}
\begin{center}
\def\arraystretch{1.1}
\begin{tabular}{|r|c|c|c|c|c|}
\hline
  $\strut\phantom{\big|}$
  $q$ (MeV/c) & $\bar{A}\,(10^{-2})$ & $\bar{B}$ & $A\,(10^{-2})$ & $B$ &
    $m_N^*(p)/m_N$ \\
\hline\hline
  300      & $-1.671$  & 0.246 & $-0.961$ & 0.146 & 0.77 \\ \hline
  500      & $-1.671$  & 0.246 & $-0.424$ & 0.116 & 0.81 \\ \hline
  800      & $-1.671$  & 0.246 & $\phantom{-}1.034$ & 0.085 & 0.85 \\ \hline
  1000     & $-1.671$  & 0.246 & $\phantom{-}1.466$ & 0.081 & 0.86 \\ \hline
\end{tabular}
\def\arraystretch{1}
\end{center}
\caption{The dimensionless parameters characterizing the biparabolic nucleon
self-energy discussed in the text (see (\protect{\ref{eq:fithole}}) and
(\protect{\ref{eq:fitparticle}})).
Note the variation of the particle self-energy with the
momentum transfer $q$, reported in the last column.
The effective mass of the particles {\em below} the Fermi surface (holes) turns
out to be $m_N^*(h)=0.67~m_N$.
 }
\label{tab:bipar}
\end{table}

Now with the self-energy (\ref{eq:fithole}-\ref{eq:fitparticle}) the solution
of eq.~(\ref{eq:se-equation}) reads
\begin{equation}
  y=\frac{\lambda}{\kappa}\frac{1}{1+2B}-\kappa+
    \frac{1}{1+2B}\frac{1}{\kappa}(a+b\beta^2),
\end{equation}
with $a=(\bar{A}-A)/2$ and $b=(\bar{B}-B)/2$, and
the constraint (\ref{eq:se-constraint}) (non-Pauli-blocked region) is
easily shown to require
\begin{equation}
  \beta\ge\left|
    {1-\sqrt{1-
    {{\strut \displaystyle 4b}\over{\strut \displaystyle(1+2B)\kappa}}
    \left[{{\strut \displaystyle 1}\over{\strut \displaystyle(1+2B)\kappa}}
    \left(\lambda+a\right)-
    \kappa\right]}\over
    {{\strut \displaystyle 2b}\over{\strut \displaystyle(1+2B)\kappa}}}\right|.
\label{eq:bipar-ineq}
\end{equation}
Then, by expanding the square root in leading order and noticing that
(\ref{eq:eff-mass}) with the self-energy (\ref{eq:fitparticle}) yields
the expression
\begin{equation}
  m_N^*=\frac{m_N}{1+2B}
\label{eq:mstar-bipar}
\end{equation}
for the particle effective mass,
we obtain for the lower limit of the momentum integration over the effective
mass in (\ref{eq:HFtilde}) the value
\begin{equation}
  \beta_0 = \frac{1}{1+2B}\left|
    \left(\lambda^*+a^*\right)\frac{1}{\kappa^*}-\kappa^*\right|
\label{eq:beta-0}
\end{equation}
in the non-Pauli-blocked region.
In (\ref{eq:beta-0}) the star reminds us that the effective mass
(\ref{eq:mstar-bipar}) should be used in the definition of the dimensionless
energy and momentum transfer variables and $a^*=(1+2B)a$.

Finally, by inserting (\ref{eq:beta-0}) in the expressions (\ref{eq:RLHFbis})
and (\ref{eq:HFtilde}) for the longitudinal response, it is immediately
apparent that one gets {\em for the HF response the same expression as for the
free response but for the replacement of the bare nucleon mass $m_N$ with
the dressed one $m_N^*$ and for the shift}
\begin{equation}
  \lambda\to\lambda^*+a^*
\label{eq:e-shift}
\end{equation}
{\em of the energy}.

Identical results can be shown to hold in the Pauli-blocked region, although in
this case, owing to the energy shift (\ref{eq:e-shift}), the response obtained
with the approximation outlined here no longer exactly vanishes as $\lambda$
goes to zero as it should. On the other hand we should not trust the physics
of the RFG in the Pauli-blocked region anyway.

The above very useful and appealing results are crucially dependent
upon whether the leading-order expansion of the square root in
(\ref{eq:bipar-ineq}) is sufficient or not in providing a good representation
for the lower limit of the momentum integral entering into the expression of
$R_L$. That this is indeed the case will be shown in the next section.

\section{The charge response in HF}
\label{sec:HF-responses}

We have calculated $R_L$ in the HF scheme with the self-energies
discussed in sect.~\ref{sec:HF} for $q=300,$ 500, 800 and 1000 MeV/c.The
results are displayed in
Fig.~\ref{fig:RL-HF} where, in addition to the charge response of the free
Fermi
gas, three cases are shown as well, corresponding to the HF evaluation of
$R_L$ with
\begin{itemize}
\item[i)]  {the true self-energy $\Sigma^{\rmmath{HF}}$;}
\item[ii)] {the self-energy fitted with a double parabola, as explained in the
previous section;}
\item[iii)]{the analytic method, which we recall amounts to {\em dressing the
nucleon mass and shifting the energy}.}
\end{itemize}

Two features are immediately apparent from the figure.
The first relates to the outstanding agreement between the results for $R_L$
obtained with the three prescriptions referred to above. The expansion of the
square root in (\ref{eq:bipar-ineq}) indeed appears to be warranted over a
large span of momentum transfers.
The second feature relates to the systematic hardening of the response
induced by the HF mean field when compared with the free response. This
appears to be substantial, is persistent as $q$
grows and, as already pointed out, stems mostly from the quadratic piece of
the Hartree self-energy of the $\sigma$ and the $\omega$.

In this connection, it is worth recalling that for local static potentials the
Hartree self-energy {\em does not} contribute to the response. It does so in
our
case because the potentials mediated by the $\sigma$ and the $\omega$ also
depend on the average between the initial and final relative momenta of the
two interacting nucleons, $P$, as seen from the non-relativistic reduction
of the Bonn potential at order $O(1/m_N)$.

To gain further insight into the impact of the HF field on the nuclear
responses and to pave the way for a similar analysis of the additional
correlations we shall consider later, we now investigate the sum rule
${\Xi}$, the mean energy $\bar{\lambda}$ and the variance $\bar{\sigma}$ of
$R_L$. These quantities should relate to the many-body
content {\em only} of the nuclear responses, which should accordingly
be disentangled beforehand from the complete charge response.  To achieve
this we follow the
pattern established in ref.~\cite{Bar94}, where it has been shown
within the context of the RFG that a {\em reduced charge
response} $S^L(\kappa,\lambda)$ can be introduced, having an
energy integral (the {\em Coulomb sum rule}) equal to one for
$\kappa\ge\eta_F$. It is defined as follows:
\begin{equation}
  S^L(\kappa,\lambda) = \frac{v_LR_L}{X'_L},
\label{eq:S-LT}
\end{equation}
where
\begin{equation}
  v_L = \left(\frac{Q^2}{q^2}\right)^2
\end{equation}
is the usual leptonic factor and
\begin{equation}
  X'_L \equiv \frac{{\cal N} X_L}{(\kappa\eta_F^3/2\xi_F)
  (\partial\psi/\partial\lambda)}\ .
\label{eq:XprimeLT}
\end{equation}
In the above, $\xi_F=\sqrt{1+\eta_F^2}-1$,
\begin{eqnarray}
  \frac{\partial\psi}{\partial\lambda} &=& \frac{\kappa}{\tau}
   \frac{\sqrt{1+\xi_F\psi^2/2}}{{\sqrt{2\xi_F}}}
   \left[ \frac{1+2\lambda+\xi_F\psi^2}{1+\lambda+\xi_F\psi^2} \right] \\
  &=& \frac{\kappa}{\eta_F\tau}\left( \frac{1+2\lambda}{1+\lambda} \right) +
   {\cal O}[\eta_F^2]
\end{eqnarray}
and
\begin{eqnarray}
  X_L &=& v_L \frac{\kappa^2}{\tau} (G^2_E + W_2\Delta)
\end{eqnarray}
with \cite{Alb88}
\begin{eqnarray}
  W_2 &=& \frac{1}{1+\tau}[G_E^2+\tau G_M^2], \\
  \Delta &=& \frac{\tau}{\kappa^2}\left[\frac{1}{3}(\varepsilon_F^2 +
   \varepsilon_F\Gamma + \Gamma^2) + \lambda(\varepsilon_F + \Gamma) +
   \lambda^2 \right] -(1+\tau)
\end{eqnarray}
and
\begin{equation}
  \Gamma\equiv
    \rmmath{max} \left\{ \left( \epsilon_F - 2\lambda\right),\
    \kappa\sqrt{1+\frac{1}{\tau}} -\lambda\right\}\, .
\end{equation}
As in (\ref{eq:XprimeLT}) ${\cal N}$ reminds us that in taking the ratio
(\ref{eq:S-LT}) a term with ${\cal N}=Z$ should be added to a
term with ${\cal N}=N$ in both the numerator and denominator.

In terms of the reduced response (\ref{eq:S-LT}), the Coulomb sum rule, the
mean
energy and the variance, namely related to the zeroth, first and second
moments of the charge response, then read as follows:
\begin{eqnarray}
  {\Xi}^L &=& \int_0^\infty\!d\omega\,S^L(\kappa,\lambda),
     \quad\quad\qquad\\
  \bar{\lambda}^L &=& \int_0^\infty\!d\omega\,\lambda\,
    S^L(\kappa,\lambda), \quad\qquad
\end{eqnarray}
and
\begin{equation}
  \bar{\sigma}^L \,=\, \sqrt{\int_0^\infty\!d\omega\,
    S^L(\kappa,\lambda)[\lambda-\bar{\lambda}^L]^2}.
\label{eq:sigmaL}
\end{equation}
In discussing our results, we recall that the HF sum rule should be identical
to
the RFG one.
This is indeed seen to occur in Fig.~\ref{fig:sumrule-TD}a, where the momentum
behaviour of ${\Xi}^L_{\rmmath{HF}}$ and ${\Xi}^L_{\rmmath{RFG}}$ is
displayed.
In Figs.~\ref{fig:sumrule-TD}b and \ref{fig:sumrule-TD}c we display the
momentum behaviour of $\bar{\lambda}_{\rmmath{HF}}$,
$\bar{\lambda}_{\rmmath{RFG}}$ and $\bar{\sigma}_{\rmmath{HF}}$,
$\bar{\sigma}_{\rmmath{RFG}}$, respectively, for the charge
response.

It clearly appears from Fig.~\ref{fig:sumrule-TD}b that up to the largest
momenta considered the HF field hardens the charge response as previously
mentioned. With respect to $\bar{\sigma}$, it can be shown for
$\kappa\ge\eta_F$ in the RFG framework that it is related to the width
$\Delta\omega=m_N(4\kappa\eta_F/\sqrt{1+4\kappa^2})$ of the response region
according to \cite{Don92}
\begin{eqnarray}
  \bar{\sigma} &=& \frac{1}{2m_N\sqrt{5}}\frac{qk_F}{\sqrt{m_N^2+q^2}}+O(k_F^2)
    \nonumber \\
               &=& \frac{1}{\sqrt{5}}\frac{\kappa\eta_F}{\sqrt{1+4\kappa^2}} +
                 O(\eta_F^2) \\
               &=& \frac{1}{4\sqrt{5}}\frac{\Delta\omega}{m_N}+O(\eta_F^2).
    \nonumber
\label{eq:sigmaRFG}
\end{eqnarray}
Remarkably, in Fig.~\ref{fig:sumrule-TD}c one observes that in spite of the
dramatic hardening of the charge response induced by the HF field,
which {\em grows} with $q$, the difference between the variance of
$R_L^{\rmmath{HF}}$ and of $R_L^{\rmmath{RFG}}$ is quite small at low
momenta ($\sim 5\%$), then it increases with momentum transfer up to
about $q=500$ MeV/c
($\sim 20\%$), to decrease again for larger momenta where the difference
stabilizes around a value of $10\%$.

Thus, notwithstanding the violation of the energy-weighted and
energy-squared weighted sum rules induced by the HF mean field, which is in
accord with a general theorem on the contributions of the Feynman diagrams
to the sum rules \cite{Tak90}, yet it appears that these violations tend
to compensate each other in arriving at $\bar{\sigma}_{\rmmath{HF}}$. It is
as if the system reacts against those interactions that attempt to change the
width of the charge response or, equivalently, the Fermi momentum $k_F$, which
in turn fixes the density for a homogeneous system.

\section{The RPA longitudinal response}
\label{sec:RPA}

In this section we explore the influence on the charge response of
correlations between nucleons acting beyond the mean field. It might be
useful in this connection to relate the ingredients used in the present work
to Feshbach's scheme \cite{Fes92}, which splits
the Hilbert space of the nuclear Hamiltonian into a $P$-sector, where the
{\em slow} nuclear motion occurs, and a $Q$-sector, where instead the {\em
fast}
motion associated with short-range correlations between nucleons takes place.
In this framework the HF field and the RPA correlations are naturally included
in the $P$-sector, whereas the $2p-2h$, $3p-3h$, {\it etc.,\/} types of
excitations
mainly belong to the $Q$-sector. Indeed the occurrence of the latter requires
a large amount of energy which is mostly supplied by the violent
nucleon-nucleon interaction at short range and short-range encounters between
nucleons are presumably comparatively rare for kinematics in the QEP region.
Therefore, our aim will be to calculate the polarization propagator $\Pi(Q)$ in
RPA on a HF basis while ignoring the $Q$-sector contributions in the present
work.

For this purpose, we start by recalling the link between $\Pi(Q)$, which for a
homogeneous system only depends upon the two variables $|\bfmath{q}|$ and
$\omega$, and the $ph$ Green's function $G^{\rmmath{ph}}$, namely
\begin{equation}
  \Pi(Q) = i \int\frac{d^4p}{(2\pi)^4} \int\frac{d^4k}{(2\pi)^4}
    G^{\rmmath{ph}}(k+Q,k;p+Q,p)
\label{eq:Pi-Gph}
\end{equation}
(spin-isospin indices are here neglected for simplicity).
Therefore, in order to get $\Pi$, one should first solve the Galitskii-Migdal
equation \cite{Gal58} for $G^{\rmmath{ph}}$ which reads in
the RPA framework
\begin{eqnarray}
 && G^{\rmmath{ph}}(k+Q,k;p+Q,p) = -G(p+Q)\,G(p)\,(2\pi)^4\delta(k-p)
    \nonumber\\
    &&\;+iG(k+Q)\,G(k)\int\frac{d^4t}{(2\pi)^4}
    \left[U^{ph,\rmmath{dir}}(Q)-U^{ph,\rmmath{exch}}(k-t)\right]
    G^{\rmmath{ph}}(t+Q,t;p+Q,p), \nonumber\\
\label{eq:Gal-Migdal}
\end{eqnarray}
where the single-particle Green's functions $G$ are meant to describe the
fermion propagation in the HF field.

Clearly, the zeroth-order iteration of (\ref{eq:Gal-Migdal}), when inserted
into (\ref{eq:Pi-Gph}), just yields the polarization propagator for the
non-interacting relativistic Fermi gas $\Pi^0(Q)$.
The first iteration of (\ref{eq:Gal-Migdal}), on the other hand, provides, via
the direct $ph$ matrix element of the interaction $U^{ph,\rmmath{dir}}$, the
first ring (Fig.~\ref{fig:RPA-diagrams}a) of the ring expansion and, via the
exchange $ph$ matrix element $U^{ph,\rmmath{exch}}$, the first
(Fig.~\ref{fig:RPA-diagrams}b) of the infinite series of exchange terms.
When the two expansions are grouped together, the fully antisymmetrized
$\Pi^{\rmmath{RPA}}$ on a HF basis is obtained. Note that the two first-order
terms of the RPA series already embody contributions going both forward (TD)
and backward (gsc) in time.

In this paper, rather then directly solving eq.~(\ref{eq:Gal-Migdal}), we
resort
to the equivalent continuous fraction representation of
$\Pi^{\rmmath{RPA}}(\bfmath{q},\omega)$ \cite{Del85,Len80,Fes92}.
A detailed analysis of the method will be given in ref.~\cite{DeP94}, where, in
particular, it will be shown how the ground-state correlations can be
incorporated in the treatment (see also ref.~\cite{Del87}). Here, to help in
understanding the present calculation, we merely recall the basic
formulas.

In ref.~\cite{Del85} it had been shown that the polarization propagator in the
TD approximation and in the first-order continuous fraction expansion is
\begin{equation}
  \Pi^{\rmmath{TD}}(\bfmath{q},\omega) \cong
    \frac{\Pi^0_R(\bfmath{q},\omega)}{1-V(\bfmath{q})\Pi^0_R(\bfmath{q},\omega)
      -[\Pi^{(1)}_R(\bfmath{q},\omega)/\Pi^0_R(\bfmath{q},\omega)]},
\label{eq:CF-TD}
\end{equation}
where the subscript $R$ means that only the retarded component of the
free propagator is included and where
\begin{equation}
  \Pi^{(1)}_R(\bfmath{q},\omega) = -\int\frac{d^4k_1}{(2\pi)^4}
    \int\frac{d^4k_2}{(2\pi)^4} G^0_R(k_1+Q)G^0_R(k_1)G^0_R(k_2+Q)G^0_R(k_2)
    V(\bfmath{k}_1-\bfmath{k}_2).
\label{eq:Pi1R}
\end{equation}
Moreover $V(\bfmath{q})$ and $V(\bfmath{k}_1 - \bfmath{k}_2)$ represent,
respectively, the direct and the exchange $ph$ matrix elements of the Bonn
potential in the appropriate spin-isospin channel.
These formulas can be generalized to encompass the
ground-state correlations as well simply by inserting into them
the full free propagator $G^0$ and $\Pi^0$ ({\it i.~e.,\/} by including both
retarded and advanced components).
Accordingly the polarization propagator will read
\begin{equation}
  \Pi^{\rmmath{RPA}}(\bfmath{q},\omega) \cong
    \frac{\Pi^0(\bfmath{q},\omega)}{1-V(\bfmath{q})\Pi^0(\bfmath{q},\omega)
      -[\Pi^{(1)}(\bfmath{q},\omega)/\Pi^0(\bfmath{q},\omega)]},
\label{eq:CF-RPA}
\end{equation}
with
\begin{equation}
  \Pi^{(1)}(\bfmath{q},\omega) = -\int\frac{d^4k_1}{(2\pi)^4}
    \int\frac{d^4k_2}{(2\pi)^4} G^0(k_1+Q)G^0(k_1)G^0(k_2+Q)G^0(k_2)
    V(\bfmath{k}_1-\bfmath{k}_2).
\label{eq:Pi1}
\end{equation}
The HF correlations should be included, in principle, using in
eq.~(\ref{eq:CF-RPA}) $\Pi^{\rmmath{HF}}$ in place of
$\Pi^0$ and in eq.~(\ref{eq:Pi1}) $G^{\rmmath{HF}}$ in place of $G^0$
(and analogously for the TD expressions).
To simplify the calculations, we shall make use of the prescription discussed
in sect.~\ref{sec:HF}, which amounts to replacing the bare nucleon mass with
the
dressed one and to shifting the energy (see the discussion above
eq.~(\ref{eq:e-shift})).

In the following we discuss the results that formulas (\ref{eq:CF-TD}) and
(\ref{eq:CF-RPA}) yield with the Bonn potential.
We split our analysis into two parts: in the first one (subsection
\ref{sec:TD})
we deal with the charge response in the TD approximation both in the free
($R_L^{\rmmath{TD,free}}$) and in the HF ($R_L^{\rmmath{TD,HF}}$) basis; in the
second one (subsection \ref{sec:gsc}) we explore the charge response in the
full
RPA, again in the free ($R_L^{\rmmath{RPA,free}}$) and in the HF
($R_L^{\rmmath{RPA,HF}}$) basis.

\subsection{Tamm-Dancoff correlations}
\label{sec:TD}

Note that in obtaining $S^L_{\rmmath{TD,free}}$ and $S^L_{\rmmath{TD,HF}}$ we
have not used the normalizing factor (\ref{eq:XprimeLT}), but rather the
``relativized'' version of it \cite{Bar94}, namely
\begin{equation}
  X'_{L,\rmmath{app}} = {\cal N}\frac{f_L^2}{\kappa\eta_F}
    \frac{1}{\partial\psi/\partial\lambda},
\label{eq:Xprimeapp}
\end{equation}
since the TD charge responses in the free and HF bases are only approximately,
although quite accurately, relativistic in the sense previously discussed in
sect.~\ref{sec:HF}.
In (\ref{eq:Xprimeapp}) $\psi$ is meant to be given by
(\ref{eq:psi-rel}) and $f_L^2$ by (\ref{eq:fL}) for the proton (${\cal N}=Z$)
and for the neutron (${\cal N}=N$), respectively.

The {\em reduced} longitudinal responses $S^L_{\rmmath{TD,free}}$ and
$S^L_{\rmmath{TD,HF}}$, like $S^L_{\rmmath{HF}}$, should yield the same sum
rule
as the non-interacting RFG \cite{Tak90}.
In Fig.~\ref{fig:sumrule-TD}a this is seen indeed to occur with a good
accuracy.
In connection with the mean energy $\bar{\lambda}$ and the variance
$\bar{\sigma}$, it is of interest to recognize in Figs.~\ref{fig:sumrule-TD}b
and c
that the TD correlations alone ($R_L^{\rmmath{TD,free}}$) {\em do not} modify
either of these two observables with respect to the predictions of the
non-interacting Fermi gas, although in principle they do not have to fulfill
the
energy-weighted and energy-squared weighted sum rules.
This outcome is reasonably clear for the variance as seen in
Figs.~\ref{fig:TD-300}--\ref{fig:TD-1000}, where $R_L$ is displayed in HF, TD
and HF-TD approximations at $q=300,$ 500, 800 and 1000 MeV/c for
$k_F=225$ MeV/c, whereas it is somewhat less evident that there is a near
identity of the mean
energy of the RFG and of the free basis TD responses at low momenta.

To provide an orientation for understanding both the moments of the
charge response and $R_L$ itself over a wide range of momentum transfers we
start by discussing the momentum dependence of the ring and exchange diagrams.
In the former case such a dependence is
mainly dictated by the direct $ph$ matrix element of the Bonn potential,
whose sole dependence is upon the momentum transfer $q$ (and this is
substantial only in the isoscalar channel --- in the isovector one the $ph$
matrix element of the $\rho$ is negligible):
\begin{eqnarray}
V(q) &=& V_0^\sigma (q) + V_0^\omega (q) \nonumber\\
  &=& 4\left\{g_\omega^2 \Gamma_\omega^2(q)\frac{1}{q^2+m_\omega^2} -
      g_\sigma^2 \Gamma_\sigma^2(q)\frac{1}{q^2+m_\sigma^2}\right.
  \nonumber \\
  & & \left.- \left({q^2\over 8 m_N^2}\right)
 \left[g_\omega^2 \Gamma_\omega^2(q)\frac{1}{q^2+m_\omega^2} +
      g_\sigma^2 \Gamma_\sigma^2(q)\frac{1}{q^2+m_\sigma^2}\right]\right\}.
\label{eq:vq}
\end{eqnarray}
This formula most clearly shows the battle going on in the ring
framework between the $\sigma$ and the $\omega$, and won by the former, as
is apparent from Fig.~\ref{fig:vq} where $V(q)$ is displayed.
There it is seen that for momenta up to about 500 MeV/c
a substantial attraction occurs that induces a softening of
$R_L^{\rmmath{ring,free}}$. This softening is evident in panel
$(b)$ of Figs.~\ref{fig:TD-300} and ~\ref{fig:TD-500}. In contrast to
this, for larger momenta $V(q)$ is no longer strong
enough to organize much collectivity in the nuclear motion and thus fails to
affect $R_L$ significantly (see panel $(b)$ of  Fig.~\ref{fig:TD-800} and
{}~\ref{fig:TD-1000}).
It should be remarked that strictly speaking formula (\ref{eq:vq}) is not valid
because the $P$-dependence of $V$ has been ignored.
On the other hand this seems to
be warranted, since in the direct (ring) channel
$\bfmath{P} = (\bfmath{k} - \bfmath{k}')/2$,
$\bfmath{k}$ and $\bfmath{k}'$ being the hole momenta of two subsequent rings.
Thus $P$ is quite small, being bound by $k_F$.

Turning to the exchange we first observe that in this channel
$\bfmath{P} = \bfmath{q}/2$ and therefore the corresponding terms in the
Bonn potential cannot now be neglected.
Moreover, the interaction lines in the exchange diagram do not depend upon
$q$. Nevertheless, for short-range interactions, the direct and the exchange
matrix elements are actually rather similar: thus the effect of the exchange
and likewise the ring diagrams is no longer perceived at large $q$, the reason
however stemming not only from the interplay among the different mesons, but
also from the competition between the isoscalar ($\tau=0$) and the isovector
($\tau=1$) contributions, as will be illustrated in the following.

In addition to the momentum dependence it is also of importance to recall the
role played by the mass of the quanta exchanged by the nucleons.
Indeed, the larger the mass,
the shorter the range of the force and the slower the decrease with momentum
of the interaction.

On the basis of the above considerations the following comments are appropriate
concerning the results displayed in the
Figs.~\ref{fig:TD-300}--\ref{fig:TD-1000}:

i)\  As already anticipated in the Introduction the forward-going ring
diagrams, contributed to only by the $\sigma$ and $\omega$
mesons, dramatically affect the charge response, however in
``opposite'' directions.
Indeed, they lead to the occurrence of isoscalar collective modes
above (the $\omega$) and below (the $\sigma$) the QEP energy range, thus
strongly depleting the charge response itself in the QEP domain.
Actually, the $\sigma$ taken alone would truly induce a phase transition in
the RFG.
However, when the $\sigma$ and the $\omega$ are considered together in the
ring framework, their effect is substantial but not dramatic, yielding
for $q\lapp500$ MeV/c an
$R_L^{\rmmath{TD,free}}$ that is quenched, flattened and softened
with respect to $R_L^{\rmmath{RFG}}$. Then these effects fade away and
at roughly 800 MeV/c
little is left of them (see the panel (b) in
Figs.~\ref{fig:TD-300}--\ref{fig:TD-1000}).
This prevailing of the $\sigma$ over the $\omega$ in the forward-going ring
diagrams is in accord with the above discussion and
clearly goes in parallel with the analogous effect occurring in the
binding energy of nuclear matter.

ii)\  {\em All} of the mesons contribute to the exchange diagrams.
Here the $\sigma$ and $\omega$ counteract the attraction and
repulsion, respectively, that they induce through the rings, thus
providing an exchange-repulsion in the case of the $\sigma$ and an
exchange-attraction in the case of the $\omega$ (in the exchange an
additional minus sign
occurs). Furthermore, an important contribution that is substantial up to
about 500 MeV/c, particularly in the isoscalar channel, is provided by the
pion.

Where the effects on the charge response are concerned, it should be observed
that at
$q=300$ MeV/c an exchange repulsion prevails. It leads to the occurrence of a
{\em dramatic peak} in $R_L^{\rmmath{exch,free}}$ (see Fig.~\ref{fig:TD-300}c)
already referred to in the Introduction.
Fig.~\ref{fig:TD-sep} is instrumental in grasping the origin of the peak.
Specifically, in Fig.~\ref{fig:TD-sep}a the total exchange response
$R_L^{\rmmath{exch,free}}$ is displayed and compared with the one obtained
using a force mediated only by the $\sigma$ and $\omega$ (or the $\sigma$,
$\omega$ and $\rho$).
The pion's role appears to be truly stunning. Without it no peak would show up
in the exchange response. Is this effect associated with the direct pionic
contribution to $R_L^{\rmmath{TD,free}}$ or with the interference that the
pion is experiencing with the other mesons?

The answer to this question can be read in Fig.~\ref{fig:TD-sep}b where the
$R_L^{\rmmath{TD,free}}$ one would get with all of the contributions,
except for the one arising from the
interference of the pion with the other mesons, is compared with
the full TD charge response in a free basis. The large difference between the
two curves is then to be ascribed entirely to the existence in the exchange
diagrams of the interferences between the pion and the other mesons.
Note however that in the same figure the contribution $\Delta R_\pi$ due to the
pion alone is also displayed and seen to be quite out of phase with that of the
interference. Accordingly, the net pionic contribution to
$R_L^{\rmmath{TD,free}}$ is no longer as substantial as in the case of the pure
exchange response. Yet it is sufficient, in spite of the presence of the large
and attractive ring term, to bring about the hardening of the TD response seen
in Fig.~\ref{fig:TD-300}d.

For $q=500$ MeV/c the peak is essentially gone, since the influence
of the $\pi$ and the $\sigma$ on the one hand and of the $\omega$
on the other more or less balance each other,
although some hardening of the response still takes place.
It is worth observing that, in contrast to the nearly exact cancellation of
their Fock contribution to the mean field, the $\sigma$ and $\omega$
{\em do}
affect the charge response via the exchange term and that this arises
mainly from the $\sigma$, which acts in the same direction as the pion
in shifting the maximum of $R_L$ to larger energies
for momenta up to about $q=500$ MeV/c.
For larger $q$  the exchange becomes insignificant. This fading away is due
not only to the
compensation of the contributions of the different mesons (but for the $\rho$,
which is very small by itself at large momenta), but also through an
isoscalar-isovector compensation as illustrated in Fig.~\ref{fig:isospin}
for $q=1$ GeV/c. In particular, there one sees that
an appreciable pionic isoscalar exchange contribution
still occurs in spite of the large momentum.

An additional comment may be made about the roles of the
$\pi$ and $\rho$ in the exchange diagram. We first recall
that the tensor force cannot act in the charge channel \cite{Alb93}.
Accordingly, the $\pi$ and $\rho$ will be felt in $R_L$ only through the
central part of the interaction they carry. For the $\pi$ in particular, it
is the short-range component (the $\delta$-component for pointlike nucleons)
which mainly comes into play, yielding, as we have already seen, a crucial
repulsion at moderate momenta (indeed the pion is strongly attractive in the
direct spin-longitudinal channel).
In the case of the $\rho$, the finite-range component of the
central interaction is stronger than for the pion: since it has an opposite
sign to the zero-range component, a net, but much weaker, attraction mediated
by the $\rho$ thus results in the exchange channel.

In summary for the exchange contributions, we find the following:
the pattern of the total TD exchange contribution to
$R_L$ is quite complex --- it turns out to be repulsive through the decisive
action of the $\pi$ that develops mainly in the isoscalar channel, which tilts
the competition between the $\omega$ and the $\sigma$ in favour of the latter
for $q$ less than about 500 MeV/c.
Because of this repulsion an ``exchange collective mode'' appears
as a marked peak on the high-energy side of the response for $q=300$ MeV/c.
For larger $q$, the peak no longer appears and the exchange contribution
becomes very small.

iii)\  We now come back to the delicate balance taking place in the TD
framework between the ring and exchange contributions. Indeed,
when evaluated in a free basis, the trend of these is to cancel one another.
The cancellation however becomes complete only around $q=800$--1000 MeV/c.
On the other hand for lower momenta,
as we have seen (see panel (d) of Fig.~\ref{fig:TD-300}), a {\em hardening}
of the charge response occurs, signaling the dominance of the exchange in
the TD correlations that arises crucially from the $\pi$ and $\sigma$
(since this is counteracting the $\omega$ contribution).

iv)\  We have previously mentioned the strength of the HF mean field
arising from the Bonn potential. A question to be asked is then: what happens
to the TD correlations when framed in the HF basis?

Here, as before, the answer should distinguish between the range of momenta
below and above 500 MeV/c. In the former instance (panel (d) of
Figs.~\ref{fig:TD-300}
and ~\ref{fig:TD-500}) a hardening, quenching and some flattening of $R_L$
is clearly apparent. As in the case of the free basis it arises from the
exchange diagrams via the $\pi$ and $\sigma$.
For $q$ above about 500 MeV/c an almost pure HF response
shows up instead, which arises because of a substantial weakening of the force
due to a compensation taking place among both the various mesons and
TD diagrams.

\subsection{Ground-state correlations}
\label{sec:gsc}

We treat the gsc in this separate section because of their subtle intertwining
with the forward-going amplitudes, which significantly affects the
interplay between ring and exchange discussed previously.

{}From a glance at Figs.~\ref{fig:RPA-300}--\ref{fig:RPA-1000} it is
immediately
apparent that the gsc strongly enhance the impact of the ring diagrams on the
charge response, at the same time damping the influence of the exchange
contributions.
More specifically, as already pointed out, the gsc go in the direction to
suppress a hardened and enhance a softened $R_L$ and thus to respect the
energy-weighted sum rule, as they are  required to do.
It is of relevance that the two effects, when they exist (namely for
$q\lapp500$ MeV/c), have a common origin.
Indeed, we have seen in Fig.~\ref{fig:vq} that the direct $ph$
interaction is always {\em negative}. On the other hand,
neglecting for a moment the exchange term, it clearly follows from
(\ref{eq:CF-RPA}) that an enhancement
of the charge response should be expected if
\begin{equation}
  \rmmath{Re}\Pi^0(\bfmath{q},\omega)\approx\frac{1}{V(\bfmath{q})}.
\label{eq:RePi0}
\end{equation}
In Fig.~\ref{fig:RePi0} the two sides of this equation are displayed versus
$\omega$ for
$q=300$ MeV/c. It is satisfying to realize that no solution of (\ref{eq:RePi0})
occurs if only the retarded component of Re$\Pi^0$ is kept, whereas a
solution exists when the advanced component (gsc) is included as well.
Hence the peak on the low-energy side of Fig.~\ref{fig:RPA-300}b and the
absence of the peak in Fig.~\ref{fig:TD-300}b are explained.
Note that the existence or absence of a solution of eq.~(\ref{eq:RePi0})
does not depend upon the basis (either free or HF).
Note also that the peak has a width, since it is embedded in the $ph$
continuum.

Furthermore, in the exchange channel the interaction turns from attractive to
repulsive. The equation to be considered in this case reads
\begin{equation}
  \rmmath{Re}\left[
    \frac{\Pi^{(1)}(\bfmath{q},\omega)}{\Pi^0(\bfmath{q},\omega)}\right]
    =1.
\label{eq:RePi1}
\end{equation}
Although eq.~(\ref{eq:RePi1}) can be solved numerically with and without the
advanced components, it is of significance to realize that to the extent that
\begin{equation}
  \Pi^{(1)}(\bfmath{q},\omega)\approx -V(\bfmath{q})
    [\Pi^0(\bfmath{q},\omega)]^2
\label{eq:Pi1appr}
\end{equation}
represents an acceptable approximation (which is the case for the present
qualitative
discussion), then instead of (\ref{eq:RePi1}) one can consider
(\ref{eq:RePi0}),
{\em except} that now the interaction is {\em positive}.
A glance at Fig.~\ref{fig:RePi0} is then sufficient for concluding that gsc can
prevent the occurrence of collective effects in the charge response
for a {\em repulsive} interaction. Indeed, this is seen to be the case
by comparing panels (c) of Figs.~\ref{fig:TD-300} and \ref{fig:RPA-300}.
Moreover, by comparing Figs.~\ref{fig:RPA-300} to \ref{fig:RPA-1000}
with the corresponding ones in the TD approximation, it is also apparent
that the gsc are no longer really operative at large momenta.
Thus, as in the TD case, in RPA we witness a vanishing of the impact of the
correlations for increasing momenta and for the same reasons, namely,
the fight between the $\sigma$ and $\omega$ in the
ring channel, the fight of the $\pi$ and $\sigma$ against the $\omega$,
plus the isoscalar-isovector cancellation in the exchange channel.

The vanishing of the gsc at large $q$ is clearly seen as well
in panel (a) of Fig.~\ref{fig:sumrule-RPA} and indeed the Coulomb sum rule
appears to be satisfied almost perfectly at $q=1$ GeV/c.
At lower values of $q$ one observes in the same figure that the gsc
violate the Coulomb sum rule, but more markedly so when they are evaluated
in a HF basis rather than in the free one.
On the other hand, from panels (b) and (c) of Fig.~\ref{fig:sumrule-RPA},
the mean energy and the variance of the response are seen to be essentially
the same as those obtained in the TD framework.

We should now comment on the critical role played by the pion in the RPA
response. Notably, in RPA a situation arises that is symmetrical, but inverse
to the one occurring in the TD. Indeed, the peak that shows up in
Fig.~\ref{fig:RPA-300}b for the ring approximation (now containing components
going both forward and backward in time) would still be present in both
$R_L^{\rmmath{RPA,free}}$ and $R_L^{\rmmath{RPA,HF}}$, except for the role
played by the pion.
This is clearly seen to happen in Fig.~\ref{fig:interference-RPA} where, as in
Fig.~\ref{fig:TD-sep}b but now in the RPA, the response one would get with all
of the contributions added to the free one,
except for the interference of the pion itself
with all of the other mesons, is displayed and compared to the full response at
$q=300$ MeV/c. A broad peak at low energy sustained by the attraction provided
by the $\sigma$ via the rings is still clearly present. In the same figure the
$\pi$--$\sigma\omega\rho$ interference contribution is also shown together with
the direct pionic contribution $\Delta R_\pi$.

In contrast with the TD case, one sees that although these two terms are still
out of phase, the dominance of the interference contribution is evident.
One can thus conclude that at moderate momenta the pion is truly
important in shaping the RPA charge response of the RFG
and that it does so by interfering with the other mesons.
It is accordingly true that the $\sigma$ and $\omega$, which are more
strongly coupled to the nucleon than the pion, rather than washing out the
impact of the latter on the charge response, are actually
effective in enhancing it.

\section{Invariance of the variance}
\label{sec:variance}

The variance of the longitudinal response, which is related to the width at
half-maximum
of the response as shown by the RFG result (\ref{eq:sigmaRFG}), is defined in
terms of the first and second moments of $R_L$. As already stressed it
represents one of the basic attributes of the charge response and, in fact,
any theory aiming to deal satisfactorily with the latter might be expected
to account for this
observable before attempting to reproduce finer details of $R_L$.

Actually when experimental data are analyzed in terms of the RFG model,
the Fermi momentum $k_F$ is usually fixed in such a way as to reproduce
the experimental width. Following this train of thought one could therefore
search for a value of $k_F$ in the HF-RPA
framework to give the same variance as that yielded by the RFG with
$k_F=225$ MeV/c,
the value that produces reasonable agreement for the width of $R_L$ when
compared with experiment for $^{12}$C at $q\lapp500$ MeV/c.  Note that
$k_F$ in the interacting model is expected to be different from the value
obtained using the free RFG.  We have performed such a search obtaining
in the HF-RPA approach the values $k_F=196$, 201, 206 and 208 MeV/c at
$q=300,$ 500, 800 and 1000 MeV/c. In so doing we have arrived at a model of the
longitudinal response that for large momenta produces the same zeroth
moment (the sum rule, see Fig.~\ref{fig:sumrule-KF}a) and, by construction, the
same second moment or variance as the RFG at any $q$.

Thus we see that the essential requirement of this procedure where
the variance is maintained is to have a lower value for $k_F$ in the
interacting problem than one has in the free RFG model. Correspondingly,
one has a lower density in the former than in the latter.  Over the
range of momentum transfers explored --- the range where the models
may be expected to be most applicable --- our results show that the
most important feature is a renormalization downwards of $k_F$ by
roughly 10\% from the free RFG value.  The residual $q$-dependence
over the range $300<q<1000$ MeV/c is quite mild.  Indeed we have no
reason to believe that it should be present at all, as its purpose
here has been simply to enforce the principle of the invariance of
the variance exactly.  When in future work we attempt to make detailed
comparisons with experiment ({\it i.e.,\/} after the missing ingredients
discussed in the Introduction have been incorporated in our model) the
starting point for application of the present interacting model to
$^{12}$C will likely be with $k_F$ fixed to around 200 MeV/c.  An
immediate consequence of this statement is that the results presented
here using $k_F=225$ MeV/c and the HF-RPA model will constitute our
starting point when discussing medium-heavy nuclei, since the corresponding
free RFG values of $k_F$ are in the range 267--252 MeV/c for $q=500$--1000
MeV/c.

Returning to Fig.~\ref{fig:sumrule-KF}, we see that the basic
global element of distinction between the HF-RPA and RFG models
accordingly rests on the first moment of the response ({\it i.~e.,\/} the
mean energy or energy-weighted sum rule).
In Fig.~\ref{fig:sumrule-KF}b $\bar{\lambda}^L$ is indeed seen to be larger in
HF-RPA than in the RFG almost uniformly in $q$ (although, by comparing with
Fig.~\ref{fig:sumrule-TD}b, the gsc are seen to
quench somewhat the difference between the correlated and free mean
energies).

To expand further on this issue we display in Fig.~\ref{fig:shift-kF}
the $q$-behaviour of the shift $\bar{\epsilon}^L$ of the maximum of the
reduced charge response $S^L(\kappa,\lambda)$ with respect to that found
for the RFG, $|Q^2|/2m_N$.
Indeed it appears that the HF-RPA shifts are somewhat lower than those
of the HF-TD model, the difference disappearing at $q\approx1$ GeV/c,
in agreement with the findings of the previous section where the gsc
contribution was shown to become insignificant somewhere in between 800 and
1000 MeV/c.  In the same figure we also display the HF-RPA and the HF-TD
predictions for $\bar{\epsilon}^L$ for fixed $k_F=225$ MeV/c. The difference
between the results in the two schemes clearly displays the effect related to
the renormalization of $k_F$ --- in fact the shift of the charge response is
seen to be reduced by at least 35$\%$, the larger reduction occurring at
the lower momenta.

The general trend seen here is an increase of $\bar{\epsilon}^L$
with $q$ up to around 800 MeV/c after which it begins to fall.
The fact that $\bar{\epsilon}^L>0$ is borne out by most
experimental measurements, although the trend with $q$ is hard to
be sure of, given the current state of the data  --- in fact little
shift is seen at all in some cases \cite{Che91}, while others \cite{Yat93} show
a $q$-behaviour that is quite compatible with the results we have
obtained at this (early) stage in our model-building.  One thing is
clear, however: the size of the shift seen in Fig.~\ref{fig:shift-kF} is
larger than
that seen in the experimental studies to date.  Clearly the ingredients that
are still missing from our analysis should be incorporated before
serious comparisons with experiemnt are made and clearly also, given
the rather confused situation that still obtains on the experimental
side, new measurements are essential.

The HF-RPA charge response is shown in Fig.~\ref{fig:RL-kF} over the usual
range of momenta both for $k_F$ fixed at 225 MeV/c (corresponding to
medium-heavy nuclei) and for $k_F$ varying with $q$ according to the
principle of the invariance of the variance (and roughly representing $^{12}$C,
see above).
The overall {\em quenching} of the interaction's
effects on the longitudinal charge response is clear, as is
its evolution with $q$. Note that the quenching of the contribution to the
longitudinal response of the individual mesons is $q$-dependent as well, an
outcome possibly related to
the different ranges of the interaction carried by the different quanta.
As an example to illustrate this point, in Fig.~\ref{fig:pi-vs-sigma}
$R_L$ is displayed for $q=300$ MeV/c along with the individual contributions
of the $\pi$ and $\sigma$.
Here the renormalization of $k_F$ is the largest and accordingly
the suppression of the interaction the strongest.
One sees that the pionic contribution to $R_L$ is significantly less affected
by the change in $k_F$ than the one arising from the $\sigma$.

\section{Conclusions}

This paper represents a further investigation (among many) of the nuclear
charge
response function. It therefore seems appropriate to comment briefly on its
motivations. In addition, while most of our basic conclusions were
anticipated in the Introduction, there are nevertheless a few that bear
amplification.

Clearly one of our motivations in the present work has been to explore $R_L$
over a wider span of momenta than is
usually done, having in mind possible future experiments to be carried out for
example at M.I.T./Bates, Mainz and CEBAF. In order to accomplish this task
fulfillment of Lorentz covariance is of major importance and indeed much
attention has been paid by us in this connection.
Furthermore, in order to provide reliable predictions for $R_L$ at momenta
up to as large as 1 GeV/c, a reliable force should be utilized.
We have chosen the Bonn potential to be representative of modern approaches
in this range of energies.
Of course in calculating the longitudinal response the harmony between currents
and forces must be preserved. Thus the predictions on $R_L$ offered in this
paper should first of all be supplemented with MEC contributions before
detailed comparisons with the available experimental data are attempted.
This is a task we are currently undertaking.

We have seen that the application of the principle of the invariance
of the variance has led us to calculate the longitudinal response
at values of $k_F$ somewhat lower than those usually considered, the
latter having been determined within the context of the free RFG model
which should be expected to yield different results than our HF-RPA model.
The consequences of this renormalization to lower values of $k_F$ and hence to
lower densities for those contributions to $R_L$
that we have neglected in the present analysis, namely the $2p-2h$
excitations, the MEC effects and effects from ladder diagrams, remain
interesting issues to be explored further. All of these scale more rapidly
by extra powers of $k_F$ than the HF, TD
and RPA contributions considered by us in the present paper. Accordingly, if we
stick to the requirement of reproducing the variance of the experimental
response within the HF--RPA scheme, their contributions to $R_L$ are likely
to be substantially suppressed.
With respect to this principle it should also be observed that were our
HF--RPA theory to be able to account for the mean energy
({\it i.~e.,\/} the first moment of the experimental response) then
it would {\em ipso facto} account as well for the second moment.
This could already bring the HF--RPA framework into rather close touch with
experimental data and, as a consequence, less room would be left for the
contribution of the MEC, $2p-2h$ and ladder diagrams.
Our starting point in future work aimed at exploring these contributions will
be
to take $k_F\cong200$ MeV/c for $^{12}$C and 225 MeV/c for medium-heavy
nuclei.
It remains to be seen what the ultimate predictions will be for the
moments of the reduced charge response (including moments higher than
two that characterize the shape of the response) and how these do or do not
agree with experiment.  Such detailed comparisons will be attempted once
the missing ingredients have been incorporated in our extended model.

Moreover, clearly even at this stage in our analysis we see that it is
very desirable to have the experimental situation concerning the
longitudinal response clarified, especially given its present status
where it is characterized by conflicting data, since only then can
definitive tests be made of our predictions.

Be it as it may, our analysis shows that, even in the limited HF--TD and
HF--RPA
schemes, the major reshaping of the charge response with respect to the
predictions of the pure RFG model arises from a rather complex intertwining
among different forward- and backward-going diagrams and among the
different mesons carrying the NN interaction.
In our view it is of significance that there appears to be good reason
to believe that, of the latter, the pion
ultimately plays the crucial role in yielding the shape and the norm of the
charge response of the correlated RFG. As illustrated in the present paper this
is accomplished through quite subtle aspects of the nuclear many-body problem.
For example, it is clear that the interference occurring between the pion and
the other mesons acquires a quite different magnitude according to whether it
occurs only in forward-going processes (TD) or in the ones going backward in
time as well (RPA). It appears that precisely these interference terms
significantly enhance the role played by the pion
in the charge response. This also suggests that the
effect of the enhancement of the weak neutral current longitudinal response
that we predicted \cite{Bar94} in the framework of a purely pionic model
will survive even in the presence of other mesons.

A relevant question to be addressed for the longitudinal response
obtained in the present paper within the context of the
RPA-HF model concerns its sensitivity to the parameters that
characterize the pieces of the Bonn potential that we have employed.
There are nine such parameters altogether, namely five coupling
constants (two for the $\rho$) and four cutoffs, {\it i.e.,\/} when the
masses are taken
as fixed which might be questionable in the case of the $\sigma$-meson.

The stability of the mean field with respect to moderate variations of
the parameters appears to be well-established: we have indeed seen that
the mean field arises mainly from the quadratic terms in the self-energy
of the $\sigma$ and $\omega$ and that these have the ``same sign''.

The situation concerning the RPA correlations is different in that there
a clearcut statement cannot be made.  In the direct channel the
competition between the $\sigma$ and $\omega$ mesons, both in forward-
and backward-going rings, leads to a cancellation of the associated
contributions which can be expected to be affected by variations of
the force parameters.  This, of course, is an effect analogous to the
one that occurs for the ground-state energy.  In the exchange channel
the $\sigma$--$\omega$ competition is altered by the action of the
$\pi$ and $\rho$, especially the former.  Therefore in this case the
sensitivity
to variation of the force parameters will stem not only from contributions
that tend to have delicate cancellations, but also from the non-linearity
in $k_F$ contained in the pionic interferences with the $\sigma$ and
$\omega$.

We are presently carrying out the task of assessing the extent of these
sensitivities to details of the input potential model, despite the
complexity of such an undertaking.

\pagebreak

\pagebreak

\section*{Figure captions}
\begin{itemize}
\item[Fig.~1]
The two Feynman diagrams contributing in first order to the
self-energy: (a) tadpole (direct), (b) oyster (exchange).
\label{fig:se-diagrams}
\item[Fig.~2]
Definition of the momenta entering the Bonn potential: accordingly, one
has $\bfmath{q}=\bfmath{k}_1-\bfmath{k}'_1=\bfmath{k}'_2-\bfmath{k}_2$ and
$\bfmath{P}=(\bfmath{k}_1-\bfmath{k}_2+\bfmath{k}'_1-\bfmath{k}'_2)/4$.
\label{fig:NN-diagram}
\item[Fig.~3]
The effective mass stemming from the Hartree self-energy of the
$\sigma$ (dashed line) and of the $\omega$ (dot-dashed line) versus the Fermi
momentum $k_F$. The solid line represents the Hartree effective mass
arising from the total $\sigma+\omega$ exchange.
\label{fig:mstar}
\item[Fig.~4]
The Fock self-energy of the $\sigma$ (dashed line) and of the $\omega$
(dot-dashed line) versus the nucleon momentum for $k_F=225$ MeV/c. Note the
flatness of the total ($\sigma+\omega$) contribution.
\label{fig:se-fock}
\item[Fig.~5]
The free (dotted) and the HF charge responses at $q=300,$ 500, 800 and
1000 MeV/c. The HF response is displayed in three versions: with the true
self-energy (\protect{\ref{eq:se-s-h}}--\protect{\ref{eq:se-o-f}}) (solid),
with the parabolic self-energy
(\protect{\ref{eq:fithole}},\protect{\ref{eq:fitparticle}})
(dashed) and with the analytic method discussed in the text (dot-dashed).
Note how close the HF curves are, but for the $q=300$ MeV/c case at very low
energy. The calculation is performed for $k_F=225$ MeV/c, which should roughly
correspond to $^{12}$C.
\label{fig:RL-HF}
\item[Fig.~6]
The sum rule ${\Xi}^L$, the mean energy $\bar{\lambda}^L$ and the
variance $\bar{\sigma}^L$ of the charge response in HF (crosses), TD
(squares) and TD on a HF basis (diamonds) compared to the free RFG
results (dashed line). Note the expected fulfillment of ${\Xi}^L$ in all
instances, the hardening of the charge response induced by the HF field but
{\em not} by the TD correlations (although, in principle, these do not have
to fulfill the energy-weighted sum rule)
and the near coincidence of the variance
of the RFG with and without TD correlations. In contrast, the HF variance is
significantly larger than the RFG one, but for the lowest momenta, although
the difference between the two is slightly reduced by the TD correlations.
\label{fig:sumrule-TD}
\item[Fig.~7]
The two Feynman diagrams contributing in first order to the
RPA series: (a) ring (direct), (b) exchange.
\label{fig:RPA-diagrams}
\item[Fig.~8]
The momentum behaviour of the direct $ph$ interaction of the Bonn
potential (the small $\rho$ meson contribution is neglected).
\label{fig:vq}
\item[Fig.~9]
The charge response in the HF and TD approximations at $q=300$ MeV/c
and $k_F=225$ MeV/c.
The dotted lines represent the free RFG response.
Moreover, we display:
in (a) $R^{\rmmath{HF}}_L$ (solid);
in (b) $R^{\rmmath{ring,free}}_L$ (dash) and $R^{\rmmath{ring,HF}}_L$ (solid);
in (c) $R^{\rmmath{exch,free}}_L$ (dash) and $R^{\rmmath{exch,HF}}_L$ (solid);
in (d) $R^{\rmmath{TD,free}}_L$ (dash) and $R^{\rmmath{TD,HF}}_L$ (solid).
\label{fig:TD-300}
\item[Fig.~10]
The charge response in the HF and TD approximations at $q=500$ MeV/c
and $k_F=225$ MeV/c.
The meaning of the lines is explained in Fig.~\protect\ref{fig:TD-300}.
\label{fig:TD-500}
\item[Fig.~11]
The charge response in the HF and TD approximations at $q=800$ MeV/c
and $k_F=225$ MeV/c.
The meaning of the lines is explained in Fig.~\protect\ref{fig:TD-300}.
\label{fig:TD-800}
\item[Fig.~12]
The charge response in the HF and TD approximations at $q=1$ GeV/c
and $k_F=225$ MeV/c.
The meaning of the lines is explained in Fig.~\protect\ref{fig:TD-300}.
\label{fig:TD-1000}
\item[Fig.~13]
The charge response in the TD approximation at $q=300$ MeV/c
and $k_F=225$ MeV/c. In (a): the free RFG response (dotted) and
$R^{\rmmath{exch,free}}_L$ with the full interaction (solid), with
$\sigma+\omega$ (dashed), with $\sigma+\omega+\rho$ (dot-dashed).
In (b): $R^{\rmmath{TD,free}}_L$ with the full interaction (solid) and with
the full interaction minus the contribution arising from the interference of
the pion with the other mesons (dashed); also shown is the contribution of the
pion alone (dotted), the global contribution of $\sigma$, $\omega$ and $\rho$
(long-dashed) and the interference of the former with the latter (dot-dashed).
\label{fig:TD-sep}
\item[Fig.~14]
$R^{\rmmath{exch,free}}_L$ (solid) and the free RFG response (dotted)
in the isoscalar (a) and isovector (b) channels at $q=1$ GeV/c and $k_F=225$
MeV/c.
Also shown are the exchange responses with only the $\pi$ (dashed) and with
$\sigma+\omega+\rho$ (dot-dashed).
\label{fig:isospin}
\item[Fig.~15]
The charge response in the HF and RPA approximations at $q=300$ MeV/c
and $k_F=225$ MeV/c.
The dotted lines represent the free RFG response.
Moreover, we display:
in (a) $R^{\rmmath{HF}}_L$ (solid);
in (b) $R^{\rmmath{ring,free}}_L$ (dash) and $R^{\rmmath{ring,HF}}_L$ (solid);
in (c) $R^{\rmmath{exch,free}}_L$ (dash) and $R^{\rmmath{exch,HF}}_L$ (solid);
in (d) $R^{\rmmath{RPA,free}}_L$ (dash) and $R^{\rmmath{RPA,HF}}_L$ (solid).
\label{fig:RPA-300}
\item[Fig.~16]
The charge response in the HF and RPA approximations at $q=500$ MeV/c
and $k_F=225$ MeV/c.
The meaning of the lines is explained in Fig.~\protect\ref{fig:RPA-300}.
\label{fig:RPA-500}
\item[Fig.~17]
The charge response in the HF and RPA approximations at $q=800$ MeV/c
and $k_F=225$ MeV/c.
The meaning of the lines is explained in Fig.~\protect\ref{fig:RPA-300}.
\label{fig:RPA-800}
\item[Fig.~18]
The charge response in the HF and RPA approximations at $q=1$ GeV/c
and $k_F=225$ MeV/c.
The meaning of the lines is explained in Fig.~\protect\ref{fig:RPA-300}.
\label{fig:RPA-1000}
\item[Fig.~19]
The solution of eq.~(\protect{\ref{eq:RePi0}}) at $q=300$ MeV/c;
the curves correspond to Re$\Pi^0_{\rmmath{TD}}$ (dashed),
Re$\Pi^0_{\rmmath{gsc}}$ (dotted) and Re$\Pi^0_{\rmmath{RPA}}$ (solid).
\label{fig:RePi0}
\item[Fig.~20]
The sum rule ${\Xi}^L$, the mean energy $\bar{\lambda}^L$ and the
variance $\bar{\sigma}^L$ of the charge response in HF (crosses), RPA
(squares) and RPA on a HF basis (diamonds) compared to the free RFG
results (dashed line).
\label{fig:sumrule-RPA}
\item[Fig.~21]
The charge response in the RPA approximations at $q=300$ MeV/c
and $k_F=225$ MeV/c. For the meaning of the lines see panel (b) of
Fig.~\protect{\ref{fig:TD-sep}}.
\label{fig:interference-RPA}
\item[Fig.~22]
The sum rule ${\Xi}^L$ and the mean energy $\bar{\lambda}^L$
of the charge response in HF-RPA (diamonds) and HF-TD (crosses) for
$k_F=196,$ 201, 206, 208 MeV/c at $q=300,$ 500, 800, 1000 MeV/c respectively.
\label{fig:sumrule-KF}
\item[Fig.~23]
The shift of the quasielastic peak of the reduced charge response
in HF-RPA (solid line) and HF-TD (dashed line). The thin curves correspond
to $k_F$=225 MeV/c, the thick curves to the variable $k_F$.
\label{fig:shift-kF}
\item[Fig.~24]
The free (dotted) and the HF-RPA (dashed) charge responses
corresponding to $k_F=225$ MeV/c are displayed as functions of the energy
transfer. The solid curves represent the HF-RPA response functions where the
values of $k_F$ are 196 , 201, 206 and 208 MeV/c for
$q=300,$ 500, 800 and 1000 MeV/c respectively.
\label{fig:RL-kF}
\item[Fig.~25]
$R_L^{\rmmath{RPA,free}}$ at $q=300$ MeV/c with the full interaction
(solid), with only the $\pi$ (dashed) and with only the $\sigma$ (dot-dashed).
In panel (a) $k_F=225$ MeV/c and in panel (b) $k_F=196$ MeV/c.
\label{fig:pi-vs-sigma}
\end{itemize}

\end{document}